\documentclass[conference,10pt]{IEEEtran}
\usepackage{amsmath,cite,amsfonts,amssymb,psfrag,amsthm,paralist}
\usepackage{graphicx}
\usepackage{epstopdf}
\usepackage{rotating}
\usepackage{dsfont}
\usepackage{color}
\usepackage{balance}
\usepackage{tikz}
\if CLASSOPTIONcompsoc
\usepackage[caption=false,font=normalsize,labelfont=sf,textfont=sf]{subfig}
\else
\usepackage[caption=false,font=footnotesize]{subfig}
\fi
\usetikzlibrary{plotmarks}
\usepackage{pgfplots}
\usetikzlibrary{calc}
\usetikzlibrary{shapes,arrows}
\usetikzlibrary{decorations.markings}
\usetikzlibrary{positioning}
\pgfplotsset{compat=1.10}
\usetikzlibrary{calc}
\usetikzlibrary{shapes,arrows}
\usetikzlibrary{decorations.markings}

\usepackage[utf8]{inputenc}
\usepackage{balance}

\DeclareFontFamily{U}{mathx}{\hyphenchar\font45}
\DeclareFontShape{U}{mathx}{m}{n}{<-> mathx10}{}
\DeclareSymbolFont{mathx}{U}{mathx}{m}{n}
\DeclareMathAccent{\widebar}{0}{mathx}{"73}

\makeatletter
\newtheorem*{rep@theorem}{\rep@title}
\newcommand{\newreptheorem}[2]{%
	\newenvironment{rep#1}[1]{%
		\def\rep@title{\Cref{##1}}%
		\begin{rep@theorem}}%
		{\end{rep@theorem}}}
\newcommand*{\textlabel}[2]{%
	\edef\@currentlabel{#1}
	\phantomsection
	#1\label{#2}
}
\makeatother

\newcommand{\Enc}{\mathsf{Enc}}
\newcommand{\Dec}{\mathsf{Dec}}

\title{Semantic Security for Indoor\\ THz-Wireless Communication}

\author{
	\IEEEauthorblockN{Rebekka Schulz\textsuperscript{1}, Onur G{\"u}nl{\"u}\textsuperscript{2}, Robert Elschner\textsuperscript{3},  Rafael F. Schaefer\textsuperscript{2},\\ Carsten Schmidt-Langhorst\textsuperscript{3}, Colja Schubert\textsuperscript{3}, and Robert F. H. Fischer\textsuperscript{1}\\[3mm]}
	\IEEEauthorblockA{\textsuperscript{1}%
		Institute of Communications Engineering, Ulm University, Germany\\
		\small \texttt{\{rebekka.schulz, robert.fischer\}@uni-ulm.de}\\[1mm]
	}
	\IEEEauthorblockA{\textsuperscript{2}%
		Chair of Communications Engineering and Security, University of Siegen, Germany\\
		\small \texttt{\{onur.guenlue, rafael.schaefer\}@uni-siegen.de}\\[1mm]
	}
	\IEEEauthorblockA{\textsuperscript{3}%
		Submarine and Core Systems Group, Fraunhofer Heinrich-Hertz-Institute, Germany\\
		\small \texttt{\{robert.elschner, carsten.schmidt-langhorst, colja.schubert\}@hhi.fraunhofer.de}
	}
	\vspace*{-1cm}
}

\IEEEoverridecommandlockouts
\IEEEpeerreviewmaketitle

\begin{document}
	\maketitle

	\begin{abstract}
		Physical-layer security (PLS) for industrial indoor terahertz (THz) wireless communication applications is considered. We use a similar model as being employed for additive white Gaussian noise (AWGN) wireless communication channels. A cell communication and a directed communication scenario are analyzed to illustrate the achievable semantic security guarantees for a wiretap channel with finite-blocklength THz-wireless communication links. We show that weakly directed transmitter (Alice) antennas, which allow cell-type communication with multiple legitimate receivers (Bobs) without adaptation of the alignment, result in large insecure regions. In the directed communication scenario, the resulting insecure regions are shown to cover a large volume of the indoor environment only if the distance between Alice and Bob is large. Thus, our results for the two selected scenarios reveal that there is a stringent trade-off between the targeted semantic security level and the number of reliably and securely accessible legitimate receivers. Furthermore, the effects of secrecy code parameters and antenna properties on the achievable semantic security levels are illustrated to show directions for possible improvements to guarantee practically-acceptable security levels with PLS methods for industrial indoor THz-wireless communication applications.
	\end{abstract}
	
	%
	
	\section{Introduction}
	\label{sec:introduction}
	
	In a digital society, flexible and reliable access to high-speed communication infrastructure is a crucial requirement. Especially communication scenarios such as autonomous driving or critical infrastructures based on Internet of Things (IoT) devices create demands for both an increased capacity in the mobile network (and in the underlying fixed fiber infrastructure) and a highly secure communication infrastructure. In addition to the classic computationally intensive cryptographic primitives in communication systems, physical-layer security (PLS) methods provide an additional resource-efficient layer of security by taking advantage of, e.g., the noise in the channel to enable secure and reliable communication.
	
	A promising possibility for increasing data capacity as well as security is the use of terahertz (THz)-wireless communication technology at high carrier frequencies ($>100$~GHz). For a higher carrier frequency, the available bandwidth that determines the available reliable transmission capacity is higher. Furthermore, a higher carrier frequency reduces the size of the required antenna elements. Thus, an increase in the carrier frequency makes compact radio systems with a large number of active antenna elements (i.e., massive multiple-input/multiple-output (MIMO) systems) possible and facilitates novel PLS-MIMO-concepts.

	Initial steps have been taken in regard to the standardization and allocation of transmission frequencies in the THz range, e.g., in the resolution 767 \cite{ITU2015} of the World Radio Congress 2015, the IEEE 802.15.3d standard \cite{IEEE}, and other standards such as IEC 62311:2019 \cite{IECStandard}, which have been updated and adapted to the expected increased use of frequencies above 100 GHz. The advantage of THz compared to lower frequency millimeter-wave radiation is the availability of continuous frequency bands and larger achievable rates \cite{akyildiz2014teranets, rappaport2019wireless}. For example, THz communication at net data rates of 100~Gb/s has been experimentally demonstrated over distances ranging from 50~cm (indoor) up to 500~m (outdoor) using monolithic-integrated electronic THz front-ends \cite{castro2020ExperimentalDemonstrations}. Depending on the scenario and link distances, compact horn or reflector antennas with appropriate antenna gains were used to efficiently compensate for the free space path losses. Although the directivity of THz radiation is significantly increased, in particular for short-range indoor scenarios, security remains an important issue especially for cases with an eavesdropper (or Eve) that has high-gain antennas and for large distances between a transmitter (or Alice) and a legitimate receiver (or Bob) \cite{ma2018security}. We illustrate the secure and insecure (or security-critical) regions in two indoor THz-wireless scenarios via secrecy maps. Secrecy maps, as defined in \cite{utkovski2019learning}, enable the visualization of insecure regions, where the presence of an eavesdropper poses a threat to secure communication.
	
    
	State-of-the-art approaches to security are based on cryptographic principles that are implemented on higher layers of the protocol stack and root on assumptions of insufficient computational capabilities of adversaries and computational hardness of certain algebraic problems. Alternatively, PLS methods enable provable security directly at the physical layer by exploiting the imperfections of the communication channel \cite{BlochBarros-2011-PhysicalLayerSecurity,PoorSchaefer-2017-WirelessPhysicalLayerSecurity}. The wiretap channel (WTC) \cite{WTC} is the basic model used for PLS, where Alice communicates with Bob and she does not want an eavesdropper to obtain a non-negligible amount of information about the transmitted message. We adapt the WTC model to indoor THz-wireless communication to illustrate cases where secure THz communication is possible. We remark that WTC encoders must have a local randomness to confuse the eavesdropper \cite{WTC}. For industrial THz-wireless communication applications, one can use the security primitives called \emph{physical unclonable functions (PUFs)} as the universal local source of randomness that can be used by an encoder \cite{benimdissertation}. PUF outputs are uniformly-random secret keys that are hardware-intrinsic and unique for each device, so using PUFs is a secure, universal, and cheap digital security solution \cite{bizimEntropyTutorial}.
	
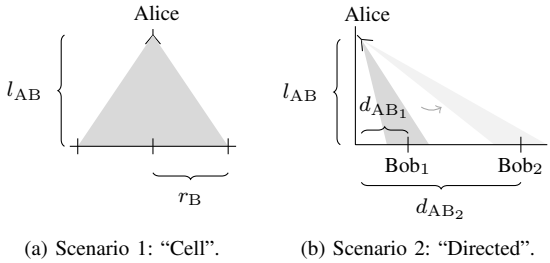
\begin{figure}
		\begin{subfloat}[Scenario~1: ``Cell''.
			\label{fig_Scenario1}]{
				\begin{tikzpicture}
\newcommand{\combineXY}[2]{#1 |- #2}
\node (bobleft) at (-1,-1.4){};
\node (bobright) at (1,-1.4) {};
\node (alice) at (0,0.08){};

\draw (alice.center) -- +(0,0.08) (alice.center)-- +(-0.1,-0.12) (alice.center) -- +(0.1,-0.12) node [black,midway,yshift=10] {\footnotesize Alice};
\fill[fill=gray!30](alice.center) --(bobleft.center) --( bobright.center);

\draw ([yshift=3] bobleft.center) -- ([yshift=-3] bobleft.center) ;
\draw ([yshift=3] bobright.center) -- ([yshift=-3] bobright.center) ;
\draw ([yshift=3] \combineXY{alice}{bobright}) -- ([yshift=-3] \combineXY{alice}{bobright}) ;
\draw [decorate, decoration={brace}] ([yshift=-10] bobright.center) -- ([yshift=-10] \combineXY{alice}{bobleft}) node [black,midway,yshift=-10] {\footnotesize $r_\mathrm{B}$};
\draw ([xshift=-3] bobleft.center) -- ([xshift=3] bobright.center);
\draw [decorate, decoration={brace}] ([xshift=-5] bobleft.center) -- ([xshift=-5]\combineXY{bobleft}{alice}) node [black,midway,xshift=-15] {\footnotesize $l_\mathrm{AB}$};
\path [decorate, decoration={brace}] (\combineXY{bobleft}{3,-1.9}) -- (\combineXY{alice}{3,-1.9}) node [black,midway,yshift=-10] {\phantom{i}};
\end{tikzpicture}
			}
		\end{subfloat}
		\begin{subfloat}[Scenario~2: ``Directed''.
			\label{fig_Scenario2}]{
				\begin{tikzpicture}
\newcommand{\combineXY}[2]{#1 |- #2}
\node (bob1) at (0.7,0){};
\node (bob2) at (2.2,0) {};
\node (alice) at (0.08,1.4){};

\draw[rotate=-45] (alice.center) -- +(-0.08,0) (alice.center)-- +(0.12,-0.1) (alice.center) -- +(0.12,0.1) node [black,midway,yshift=10] {\footnotesize Alice};
\fill[fill=gray!30](alice.center) --([xshift=-8] bob1.center) --([xshift=8] bob1.center);
\fill[fill=gray!10](alice.center) --([xshift=-10] bob2.center) --([xshift=10] bob2.center);

\draw[->,gray!70] ([xshift=5]\combineXY{bob1}{0,0.5}) to [out=-30,in=-150] ([xshift=-30]\combineXY{ bob2}{0,0.5});
\draw ([yshift=3] bob1.center) -- ([yshift=-3] bob1.center) node [black,midway,yshift=-8] {\footnotesize Bob$_1$};
\draw [decorate, decoration={brace}] (\combineXY{bob2}{3,-0.5}) -- (\combineXY{alice}{3,-0.5}) node [black,midway,yshift=-10] {\footnotesize $d_\mathrm{AB_2}$};
\draw [decorate, decoration={brace}]  (\combineXY{alice}{3,0.15})-- (\combineXY{bob1}{3,0.15}) node [black,midway,yshift=10] {\footnotesize $d_\mathrm{AB_1}$};
\draw ([yshift=3] bob2.center) -- ([yshift=-3] bob2.center) node [black,midway,yshift=-8] {\footnotesize Bob$_2$};
\draw (0,0) -- ([yshift=5]\combineXY{0,0}{alice}) (0,0) -- ([xshift=10]\combineXY{bob2}{0,0});
\draw [decorate, decoration={brace}] (-0.2,0) -- (\combineXY{-0.2,0}{alice}) node [black,midway,xshift=-15] {\footnotesize $l_\mathrm{AB}$};
\end{tikzpicture}
			}
		\end{subfloat}
		\caption{THz communication scenarios. Alice is placed in an elevated place and transmits messages to Bob who is placed on the floor. The height difference between them is denoted as $l_\mathrm{AB}$.}
		\vspace*{-0.7cm}
	\end{figure}

	We consider two typical communication scenarios for THz communication in a large room, such as a factory hall. In Scenario~1, defined as ``cell communication'', a weakly directed transmitter communicates with multiple legitimate receivers positioned within a circle on the floor defined by the 3~dB power cone beneath the transmitter. Scenario~1 is depicted in Fig. \ref{fig_Scenario1}. Furthermore, for Scenario~2, defined as ``directed communication'', a transmitter with antennas that are more directed and that can be adaptively aligned communicates with legitimate receivers at varying distances, as depicted in Fig. \ref{fig_Scenario2}. We evaluate the semantic security levels for these scenarios as a function of the eavesdropper's position to characterize secure and insecure regions in the room. If one can prevent the physical access of unknown devices/individuals to insecure regions, semantic security can be guaranteed. For an increasing distance between Alice and Bob, the insecure regions are shown to be enlarged. Moreover, if secure communication with multiple legitimate receivers is desired for fixed transmit antenna alignments, we show that there is a stringent trade-off between the volume of the insecure regions and the possible legitimate receiver positions as a function of the antenna directivity.
        
	\section{System Model}\label{sec:systemmodel}

	Consider a WTC, where the channels from Alice to both Bob and Eve are THz wireless links so that the symbols $X^n = (X_1, X_2, \dots, X_n)$ are transmitted over a discrete-time memoryless wireless THz channel to Bob (or Eve) who observes symbols $Y^n$ (or who overhears symbols $Z^n$). In the equivalent complex baseband domain, the main channel, i.e., the channel between Alice and Bob, is modeled as
	\begin{equation}
	Y_i = H_{\mathrm{AB},i}X_i + N_{\mathrm{AB},i} \label{eq:WTCBob}
	\end{equation}
	for the time index $i=1,2,\ldots,n$, where $H_{\mathrm{AB},i}$ is the channel coefficient and $N_{\mathrm{AB},i}$ is the independent complex AWGN component for the main channel. Similarly, the channel between Alice and Eve is modeled as
	\begin{equation}
	Z_i = H_{\mathrm{AE},i}X_i + N_{\mathrm{AE},i}\label{eq:WTCEve}
	\end{equation}
	for $i=1,2,\ldots,n$, where $H_{\mathrm{AE},i}$ is the channel coefficient and $N_{\mathrm{AE},i}$ is the independent complex AWGN component on the channel between Alice and Eve.

	In a typical indoor THz-wireless communication setup, multiple rays travel along different paths from a transmitter to a receiver and experience different attenuations and delays \cite{han2014multi}. Because of different angles of departure and arrival, the paths experience different gains, dependent on the antenna alignment and directivity. Furthermore, depending on the carrier frequency $f_\mathrm{c}$ and path length $d$, each ray experiences a free-space path loss according to \cite{friis1946note}
	\begin{equation}
	H_{\mathrm{fspl}}(f_\mathrm{c},d) = \left(\frac{\mathrm{c}_0}{4\pi f_\mathrm{c} d}\right)^2
	\end{equation}
	where $\mathrm{c}_0$ is the speed of light in vacuum. In general, reflected rays suffer from a larger free-space path loss due to an increased path length compared to the line-of-sight (LOS) communication. Moreover, reflection loss causes further attenuation of the signal \cite{han2014multi,piesiewicz2007scattering}. For THz communication, many common building materials cannot be considered as smooth surfaces, thus, the loss due to rough surface scattering has to be also considered \cite{piesiewicz2007scattering}. If the antennas are aligned along the LOS path, the reflected paths experience significantly smaller gains for THz-wireless communication due to the angle differences \cite{priebe2012impact}, which significantly increases the differences in signal strengths unlike for wireless communication. Thus, the reflected paths are negligible in our scenarios and we consider only the dominating LOS path. The received signal power of Bob $P_\mathrm{B}$ and Eve $P_\mathrm{E}$ are then~\cite{friis1946note}
	\begin{align}
	&P_\mathrm{B} = P_\mathrm{A}G_\mathrm{A}G_{\mathrm{B}}H_{\mathrm{fspl}}(f_\mathrm{c},d_{\mathrm{A}\mathrm{B}}) ,\label{eq:ReceivedPowerBob}\\
	&P_\mathrm{E} = P_\mathrm{A}G_\mathrm{A}G_{\mathrm{E}}H_{\mathrm{fspl}}(f_\mathrm{c},d_{\mathrm{A}\mathrm{E}})\label{eq:ReceivedPowerEve}
	\end{align}
	where $P_\mathrm{A}$ is the average transmit power, $G_\mathrm{A}$, $G_\mathrm{B}$, and $G_\mathrm{E}$ are the respective antenna gains for Alice, Bob, and Eve, and $d_{\mathrm{A}\mathrm{B}}$ and $d_{\mathrm{A}\mathrm{E}}$ are the path lengths from Alice to Bob and Eve, respectively. 
	The channel coefficients in (\ref{eq:WTCBob}) and (\ref{eq:WTCEve}) include the attenuation experienced during the propagation and are given by $\|H_{\mathrm{AB}}\| = \sqrt{P_B/P_A}$ and $\|H_{\mathrm{AE}}\| = \sqrt{P_E/P_A}$.
	
	At the receiver, in addition to the attenuation during the propagation across the room, thermal noise is added with noise power \cite{anderson2006digital}
	\begin{equation}\label{eq:Noise}
	N = \mathrm{k}_\mathrm{B}TBF
	\end{equation} 
	where $\mathrm{k}_\mathrm{B}$ is the Boltzmann constant, $T$ is the temperature in Kelvin, and $B$ the is transmission bandwidth. For the resulting signal-to-noise ratio (SNR) calculations, also the corresponding noise factor $F$ of the receiver front-end needs to be considered. This noise component is assumed to be distributed according to a white Gaussian distribution; thus, we use well-known expressions given for AWGN channels, as in, e.g., \cite{utkovski2019learning}. Hence, the noise components $N_{\mathrm{AB},i}$ in (\ref{eq:WTCBob}) and $N_{\mathrm{AE},i}$ in (\ref{eq:WTCEve}) are AWGN with noise variances $\sigma_{N_\mathrm{AB}}^2 = \sigma_{N_\mathrm{AE}}^2 = N$. The SNR experienced at Bob is then calculated using the power of the received signal and the noise power, $\mathrm{SNR}_\mathrm{AB} =P_\mathrm{B}/N.$
	Similarly, the eavesdropper's SNR is $\mathrm{SNR}_\mathrm{AE} = P_\mathrm{E}/N\label{eq:SNREve}$.

	\section{Semantic Security for THz Communication}\label{sec:semanticsec}

	Rooting on information-theoretic principles, there have been multiple secrecy metrics defined, including weak and strong secrecy. In this work, we consider the \emph{semantic security} that provides the strongest secrecy guarantees; see, e.g., \cite{Bellare}. In a semantically secure scheme, a computationally unbounded eavesdropper cannot gain a non-negligible amount of information about the message. We use the semantic security definition and results from \cite[Section~II]{utkovski2019learning} given for wireless AWGN channels, and summarize them below. Using this approach, a bound on the leaked amount of information can be given for the used code and channel characteristics.

	We first define the main parameters and functions used. $D_{\alpha}(\cdot\|\cdot)$ denotes the R{\'e}nyi divergence of order $\alpha\in\mathbb{R}^+\setminus \{1\}$, $C_{\mathrm{AB}}$ and $C_{\mathrm{AE}}$ denote the capacity of the channel from Alice to Bob and Eve, respectively, $R$ is the secrecy (or wiretap) code rate, and $L$ is the local randomness rate of the secrecy code that is used to confuse Eve. Define also the target average probability of error $\varphi\in[0,1]$ at Bob guaranteed for all possible positions of Bob. Similarly, denote the target semantic security level as $\delta\in[0,1]$.
	
	Denote the probability distribution of the transmitted message $M$ as $P_M$. $\Enc(\cdot)$ is an encoding function with a possible realization $g$ and $\Dec(\cdot)$ is a decoding function. Then, the resulting semantic security level $\delta_{E}\in[0,1]$ is \cite{utkovski2019learning}
	\begin{equation*}
	\max_{\substack{P_M,\\\Enc(\cdot)}}\left(\!\max_{\Dec(\cdot)}\Pr[\Dec(Z^n)\!=\!\Enc(M)] - \max_g \Pr[g\!=\!\Enc(M)]\!\right)\!\!.
	\end{equation*}
	It can be shown that a target semantic security level of $\delta$ implies that the average probability of error at Eve is
	\begin{equation}
	\bar{e}_\mathrm{E} \geq 1 - \delta - 1/2^b\label{eq:Eveerrorsemsecbit}
	\end{equation}
	when Eve tries to reconstruct only $b$ bits of information from the message $M$ since (\ref{eq:Eveerrorsemsecbit}) can be equivalently written as $b\leq -\log_2(1-\delta-\bar{e}_{\mathrm{E}})$ \cite{utkovski2019learning}. This implies that a smaller target value of $\delta$ corresponds to a higher security. Similarly, define the resulting average probability of error at Bob as $\bar{e}_{\mathrm{B}}$.

    We next summarize the fundamental semantic security results used. Suppose there exists a set of parameters $\alpha_1\in (0,1)$, $\alpha_2>1$, and $\lambda_1,\lambda_2,\epsilon_1,\epsilon_2>0$ that satisfy
	\begin{align}
	&\varphi = \mathrm{e}^{-n(1-\alpha_1)(D_{\alpha_1}(P_{\mathrm{XY}}\|P_\mathrm{X}P_\mathrm{Y})-C_{\mathrm{AB}}+\lambda_1)}\nonumber\\
	&\qquad\qquad+ 2^{-n(C_{\mathrm{AB}}-R-L-\lambda_1)}+\epsilon_1, \label{eq:reliability}\\
	&\delta = \mathrm{e}^{-n(1-\alpha_2)(D_{\alpha_2}(P_\mathrm{XZ}\|P_\mathrm{X}P_\mathrm{Z})-C_{\mathrm{AE}}-\lambda_2)}\nonumber\\
	&\qquad\qquad+ 2^{-n(L-C_{\mathrm{AE}}-\lambda_2)/2}+2\epsilon_2. \label{eq:securitylevel}
	\end{align}	
	 Then, the probability $\Pr[\{\bar{e}_{\mathrm{B}}>\varphi\}\cup\{\delta_{\mathrm{E}}>\delta\}]$ of the union of unwanted events can be upper bounded by a term that vanishes doubly-exponentially in the secrecy code blocklength $n$ \cite{IgorResolvability,utkovski2019learning}, which shows the existence of an encoder-decoder pair that satisfies the target reliability-security pair $(\varphi,\delta)$ jointly.
	
	Since the optimal input and outputs are Gaussian random variables for the THz-wireless channels defined above, we can use the closed-form representation of the R{\'e}nyi divergences for bivariate Gaussian distributions in (\ref{eq:reliability}) and (\ref{eq:securitylevel}). Suppose $f_i$ and $f_j$ are bivariate Gaussian distributions with correlation coefficients $\rho_i$ and $\rho_j$, respectively. Then, the R{\'e}nyi divergence $D_\alpha(f_i\|f_j)$ of order $\alpha$ is given as \cite[pp. 72]{gil2011renyi}
	\begin{align*}
	\!\frac{1}{2} \ln \left(\frac{1\!-\!\rho_j^2}{1\!-\!\rho_i^2}\right)-\frac{1}{2(\alpha\!-\!1)}\ln\left(\frac{1-{(\alpha\rho_j+(1-\alpha)\rho_i)}^2}{(1\!-\!\rho_j^2)}\right).
	\end{align*}
	In (\ref{eq:reliability}) and (\ref{eq:securitylevel}), $\rho_i$ corresponds to the correlation coefficient between the input and output of the respective channels and $\rho_j= 0$. Similarly, the channel capacities are calculated by using the SNR of the complex AWGN channels as
	\begin{align}
	&C_{\mathrm{AB}} = \log_2(\mathrm{1\!+\!SNR}_{\mathrm{AB}}),\quad C_{\mathrm{AE}} = \log_2(\mathrm{1\!+\!SNR}_{\mathrm{AE}}).
	\end{align}

	For comparisons below, we fix the blocklength $n$ and secrecy code rate $R$ that corresponds to the code rate if there is a secrecy constraint on the transmitted messages. In Scenario~1, we choose the transmit power $P_{\mathrm{A}}$ and $L$ such that the minimum possible value of $\varphi$ over all possible $\alpha_1$ and $\lambda_1$ according to (\ref{eq:reliability}) is equal to the target value. The solution is not unique, so the values of $P_\mathrm{A}$ and $L$ have to be chosen jointly, which allows imposing practical limits on them. Suppose the higher-order terms $\epsilon_1$ and $\epsilon_2$ are negligible, so they are not included in the optimizations. Since all parameters are fixed for the transmission to multiple locations within the chosen circle, the parameters are selected such that the target value $\varphi$ is achieved if Bob is placed on the edge of the circle. For positions within the circle, higher reliability is obtained due to smaller path losses and larger antenna gains. For a fixed environment, the channel parameters are assumed to stay constant over multiple transmissions.  
	
	For Scenario~2, the approach is similar except that $L$ is varied depending on the legitimate receiver position such that $\varphi$ is guaranteed for Bob. Thus, in this scenario, $L$ is not fixed for multiple transmissions but has to be adapted according to the receiver position. The transmit power is chosen such that for different choices of $L$, the target reliability $\varphi$ is achieved for all desired legitimate receiver positions. For both scenarios, the target security level $\delta$ can then be obtained by minimizing (\ref{eq:securitylevel}) over all possible $\alpha_2$ and $\lambda_2$. Following this approach, we obtain the target semantic security level $\delta$ for each eavesdropper position separately, the results of which serve as an outer (or converse) bound for the achievable semantic security levels. We depict the security levels for Scenarios 1 and 2 for a large set of possible eavesdropper positions in a two-dimensional color map and refer to this as a \emph{secrecy map}, as proposed in \cite{utkovski2019learning}. Secrecy maps show the semantic security levels that can be achieved for the chosen communication scenario if an eavesdropper is present in the corresponding location. Thus, they can be used to evaluate the insecure regions for different communication scenarios, e.g., to assess the impact of various room properties, antenna gains, and code parameters.

	\section{Secure THz Communication Results and Comparisons}\label{sec:THZresults}

	Consider Scenarios 1 and 2 defined above. For both scenarios, assume without loss of generality that the antenna gains for Alice and Bob are the same, i.e., $G_\mathrm{A}=G_\mathrm{B}$. Furthermore, we assume that all receivers share the same noise figure, which is equal to $9$~dB. The temperature is $T = 290$~K, the bandwidth $B = 1$~GHz, and the carrier frequency $f_\mathrm{c} = 300$~GHz. 
	
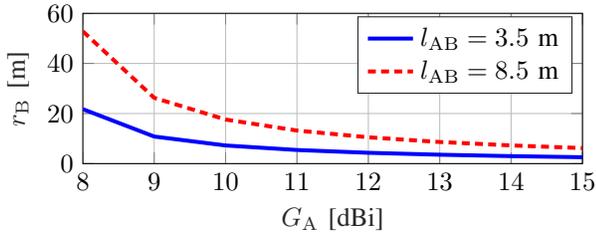
\begin{figure}
		\centering
%
%
\begin{tikzpicture}

\begin{axis}[%
width = 0.75\linewidth,
height = 2.0cm,
scale only axis,
xmin=8,
xmax=15,
xlabel style={font=\color{white!15!black}},
xlabel={$G_{\mathrm{A}}$  [dBi]},
ymin=0,
ymax=60,
ylabel style={font=\color{white!15!black}},
ylabel={$r_{\mathrm{B}}$ [m]},
axis background/.style={fill=white},
grid = both,
legend style={legend cell align=left, align=left, draw=white!15!black}
]
\addplot [color=blue,line width=1.5pt]
table[row sep=crcr]{%
	8	21.7512801896014\\
	9	10.8139800462475\\
	10	7.2492974146497\\
	11	5.44004660529446\\
	12	4.32499615676616\\
	13	3.55793416550677\\
	14	2.9919713528567\\
	15	2.55396025505452\\
};
\addlegendentry{$l_{\mathrm{AB}} = 3.5$~m}

\addplot [color=red,line width=1.5pt, densely dashed]
table[row sep=crcr]{%
	8	52.8245376033176\\
	9	26.2625229694581\\
	10	17.605436578435\\
	11	13.2115417557151\\
	12	10.5035620950035\\
	13	8.64069725908788\\
	14	7.26621614265198\\
	15	6.20247490513239\\
};

\addlegendentry{$l_{\mathrm{AB}} = 8.5$~m}

\end{axis}
\end{tikzpicture}%
		\vspace*{-0.3cm}
		\caption{Radius $r_\mathrm{B}$ of the circle on the floor that is defined by the 3 dB power cone that depends on the transmit antenna gain $G_\mathrm{A}$ for two height differences $l_{\mathrm{AB}}$ between the transmitter and legitimate receiver.}
		\label{fig_Scenario1_RadiusBob}
		\vspace*{-0.5cm}
	\end{figure}

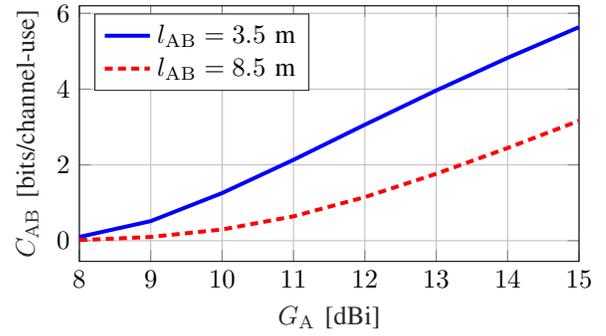
\begin{figure}
		\centering
%
%
\begin{tikzpicture}

\begin{axis}[%
width = 0.75\linewidth,
height = 3.4cm,
scale only axis,
xmin=8,
xmax=15,
xlabel style={font=\color{white!15!black}},
xlabel={$G_{\mathrm{A}}$ [dBi]},
ylabel style={font=\color{white!15!black}},
ylabel={$C_{\mathrm{AB}}$ [bits/channel-use]},
axis background/.style={fill=white},
legend pos = north west,
grid = both,
legend style={legend cell align=left, align=left, draw=white!15!black}
]
\addplot [color=blue,line width=1.5pt]
  table[row sep=crcr]{%
8	0.0992861205453422\\
9	0.517054386103492\\
10	1.25713473468916\\
11	2.13379292109932\\
12	3.06064325780953\\
13	3.96469166613948\\
14	4.82476375792439\\
15	5.63560477749949\\
};
\addlegendentry{$l_{\mathrm{AB}} = 3.5$~m}

\addplot [color=red,line width=1.5pt, densely dashed]
  table[row sep=crcr]{%
8	0.0163506450063334\\
9	0.0990846763206207\\
10	0.295566074213861\\
11	0.642870604636716\\
12	1.1502906252872\\
13	1.76567789603005\\
14	2.4458891495156\\
15	3.16503457898478\\
};
\addlegendentry{$l_{\mathrm{AB}} = 8.5$~m}

\end{axis}
\end{tikzpicture}%
			\vspace*{-0.3cm}
		\caption{Minimum channel capacity $C_{\mathrm{AB}}$ for Scenario 1 when the legitimate receiver is located within the 3 dB power cone that depends on the transmit antenna gain $G_\mathrm{A}$ for two height differences $l_{\mathrm{AB}}$ between the transmitter and legitimate receiver.}
		\label{fig_Scenario1_Capacity}
		\vspace*{-0.3cm}
	\end{figure}

	For Scenario 1 depicted in Fig.~\ref{fig_Scenario1}, a weakly directed transmitter is mounted on the ceiling facing downwards. The legitimate receiver is positioned on the floor below the transmitter within a circle with radius $r_{\mathrm{B}}$. The size of the circle depends on the transmit antenna directivity and the height $l_{\mathrm{AB}}$ of the room such that the receiver stays always within the 3~dB power cone. In Fig.~\ref{fig_Scenario1_RadiusBob}, the radius of this circle is depicted for multiple transmit antenna gains $G_\mathrm{A}$ and height differences $l_{\mathrm{AB}}$ between the transmitter and receiver. For Fig.~\ref{fig_Scenario1_RadiusBob}, we assume that the transmitter is placed $0.5$~m below the ceiling and the receivers have a height of $1$~m. For large halls with heights of $5$~m and $10$~m, this corresponds to the chosen height differences of $3.5$~m and $8.5$~m, respectively.
	With an increasing transmit antenna gain, the radius of the circle, thus, also the allowed region size for legitimate receiver positions shrinks. A smaller height difference between the transmitter and receiver also results in a smaller radius. Furthermore, in Fig.~\ref{fig_Scenario1_Capacity} the minimum channel capacity for a legitimate receiver is depicted. For Fig.~\ref{fig_Scenario1_Capacity}, we consider the same transmit power of $9$~mW for all used antenna gains. The minimum capacity is experienced by a receiver positioned on the edge of the circle since this position corresponds to the maximum path length and the minimum transmit antenna gain. With an increasing antenna gain, the circle becomes smaller, hence, the maximum possible path length to a legitimate receiver also decreases, which results in a smaller path loss. Furthermore, with an increasing antenna gain, the received signal power increases. Both aspects contribute to the increasing channel capacity because of an increasing antenna gain. Thus, with weaker directed antennas more legitimate receivers can be communicated with, but the minimum channel capacity decreases. 
	
\begin{figure}
		\centering
%
%
\definecolor{mycolor1}{rgb}{0.00000,0.44700,0.74100}%
\definecolor{mycolor2}{rgb}{0.85000,0.32500,0.09800}%
\definecolor{mycolor3}{rgb}{0.92900,0.69400,0.12500}%
\begin{tikzpicture}

\begin{axis}[%
width = 0.75\linewidth,
height = 3.2cm,
scale only axis,
xmin=10,
xmax=30,
xlabel style={font=\color{white!15!black}},
xlabel={$G_{\mathrm{A}}$ [dBi]},
ylabel style={font=\color{white!15!black}},
ylabel={$C_{\mathrm{AB}}$ [bits/channel-use]},
axis background/.style={fill=white},
grid = both,
legend pos = north west,
legend style={legend cell align=left, align=left, draw=white!15!black}
]
\addplot [color=mycolor1,line width=1.5pt]
table[row sep=crcr]{%
10	0.15053482461738\\
15	1.07025796269279\\
20	3.58473260377326\\
25	6.79416732750343\\
30	10.1043481828886\\
};
\addlegendentry{$d_{\mathrm{AB}} = 5$~m}

\addplot [color=mycolor2,line width=1.5pt, dashdotted]
table[row sep=crcr]{%
10	0.0474676794482948\\
15	0.41629225012913\\
20	2.11933450750397\\
25	5.10640474710071\\
30	8.39014089231306\\
};
\addlegendentry{$d_{\mathrm{AB}} = 15$~m}

\addplot [color=black,line width=1.5pt, dotted]
table[row sep=crcr]{%
10	0.0197983902113381\\
15	0.186726118470753\\
20	1.25203977678343\\
25	3.88925874435096\\
30	7.12078671386709\\
};
\addlegendentry{$d_{\mathrm{AB}} = 25$~m}

\end{axis}
\end{tikzpicture}%
			\vspace*{-0.3cm}
		\caption{Channel capacity $C_{\mathrm{AB}}$ for Scenario 2 for different antenna gains $G_\mathrm{A}$. The distance between the transmitter and legitimate receiver $d_{\mathrm{AB}}$ is measured horizontally on the floor.}
		\label{fig_Scenario2_Capacity}
		\vspace*{-0.48cm}
	\end{figure}
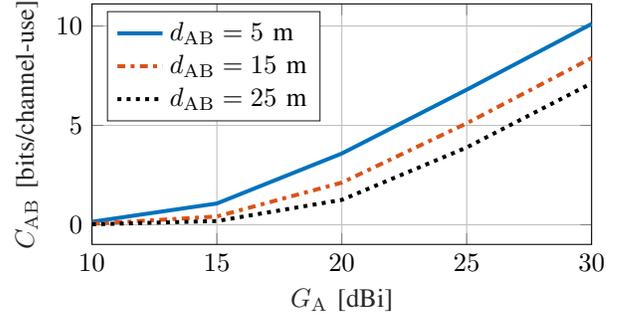

		In Scenario 2 depicted in Fig.~\ref{fig_Scenario2}, the transmitter is placed on the wall on one side of the room with a height difference of $l_{\mathrm{AB}} = 8.5$~m to the legitimate receiver. Compared to Scenario~1, strongly directed antennas are assumed in this scenario. The transmit antenna alignment is not fixed but can be adapted depending on the location of the legitimate receiver. Fig.~\ref{fig_Scenario2_Capacity} shows the channel capacity for this scenario, which depends on the used antennas for three different horizontal distances $d_{\mathrm{AB}}$ between the transmitter and the legitimate receiver. For all receiver positions, a transmit power of $0.5$~mW is used. Similar to Scenario~1, the channel capacity increases with stronger directivity and smaller distances. High gain antennas are necessary to compensate for the pathloss caused by large distances to provide a large channel capacity.
		
\begin{figure*}
		\centering
		\begin{subfloat}[Scenario~1: Alice is in the middle and Bob is within the smaller circle around the center with a height difference of $l_\mathrm{AB} = 3.5$~m. Alice, Bob, and Eve use antennas with antenna gains $G_\mathrm{A} =G_\mathrm{B}= G_\mathrm{E} = 10$~dBi. \label{fig_Scenario1_SecrecyMap_GE10}]{
				\includegraphics[width=0.35\textwidth, height=0.32\textheight, keepaspectratio]{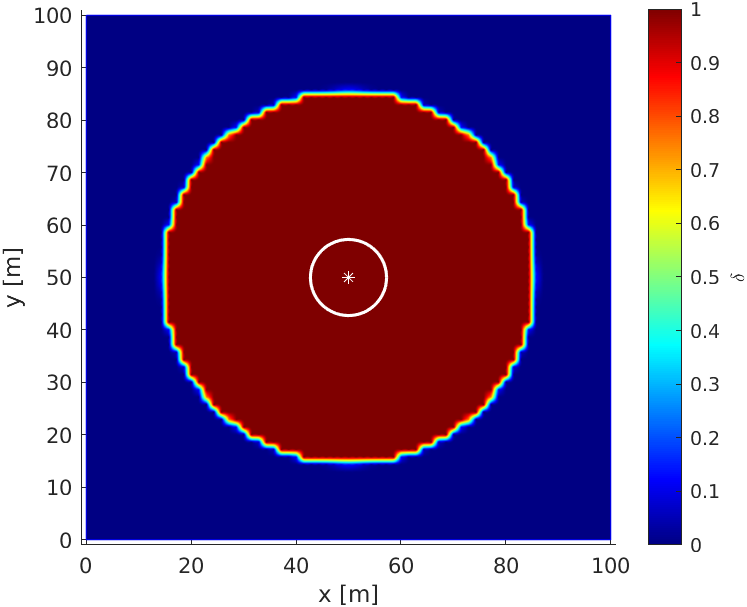}}
		\end{subfloat}
		\hspace*{1.6cm}
		\begin{subfloat}[Scenario~2: Alice is on the wall $l_\mathrm{AB}=8.5$~m above Bob with a horizontal distance of $d_\mathrm{AB}=15$~m. Alice and Bob use antennas with gains $G_\mathrm{A} =G_\mathrm{B}= 20$~dBi, whereas Eve uses a strongly directed antenna with gain $G_\mathrm{E} = 25$~dBi. 
			\label{fig_Scenario2_SecrecyMap_dist15}]{
				\includegraphics[width=0.35\textwidth, height=0.32\textheight, keepaspectratio]{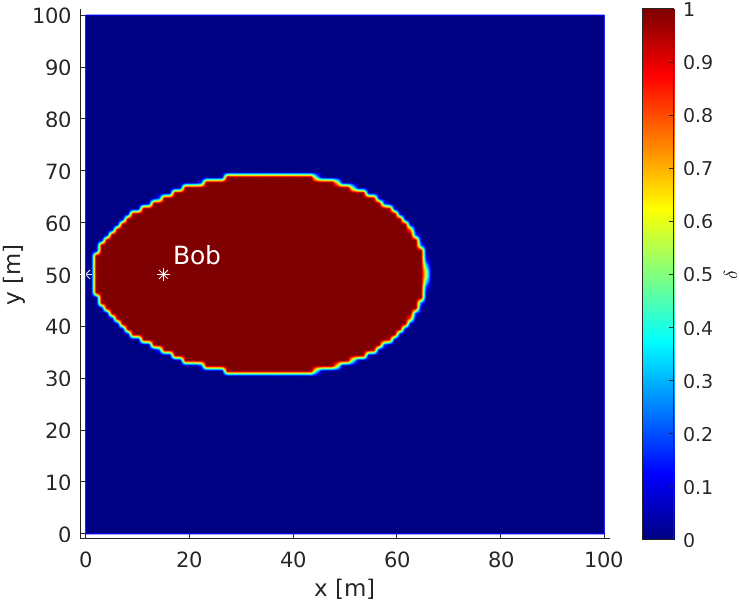}
			}
        \end{subfloat}
        \vspace*{-0.0cm}
		\caption{Secrecy Maps for the chosen scenarios. Semantic security levels are depicted for the corresponding eavesdropper positions. Pre-determined parameters for evaluations: $n = 2000$, $\varphi \leq 10^{-3}$, and $R = 0.2$ bit/channel-use.}
		\vspace*{-0.5cm}
	\end{figure*}

	In Fig.~\ref{fig_Scenario1_SecrecyMap_GE10}, a secrecy map for Scenario~1 is depicted, where we mainly follow the approach in \cite{utkovski2019learning} to plot the secrecy map. Alice, Bob, and Eve use antennas with gains of $G_\mathrm{A} = G_\mathrm{B} = G_\mathrm{E} = 10$~dBi. The height difference between transmitter and legitimate receivers are $l_\mathrm{AB} = 3.5$~m. Bob is located within the smaller depicted circle with radius $r_{\mathrm{B}} = 7.2$~m. 

	For an appropriately chosen secrecy code, the probability that the average error at Bob is greater than $\varphi$ or that the semantic security level achieved for the transmitted messages is greater than $\delta$ is doubly-exponentially small; see Section~\ref{sec:semanticsec}. We chose the secrecy code parameters such that $\varphi \leq 10^{-3}$ for all legitimate receiver locations within the circle with radius $r_\mathrm{B}$. As shown in Fig.~\ref{fig_Scenario1_RadiusBob}, $r_\mathrm{B}$ depends on the height difference between the transmitter and the receivers as well as the antenna gains. Using the same secrecy code parameters, the semantic security level $\delta$ is calculated for different eavesdropper locations. A color map is used to depict the achieved semantic security level $\delta$ for the corresponding eavesdropper positions. A higher semantic security, corresponding to a smaller value of $\delta$, can be achieved if the presence of an eavesdropper in a second circle with radius $r_\mathrm{E}$ can be prevented. The size of this circle depends on the antenna gains of the transmitter and eavesdropper, the secrecy code parameters, and the height difference between the transmitter and the legitimate receiver. For a strongly directed antenna available at an eavesdropper, the insecure region enlarges. For the scenario shown in Fig.~\ref{fig_Scenario1_SecrecyMap_GE10}, where Alice, Bob, and Eve use antennas with the same gain $G_\mathrm{A} = G_\mathrm{B} = G_\mathrm{E} = 10$~dBi, the radius is $r_\mathrm{E} = 21.1$~m.
	
	In Fig.~\ref{fig_Scenario2_SecrecyMap_dist15} the secrecy map for Scenario 2 is shown, where the horizontally measured distance from the legitimate receiver to the transmitter is $d_\mathrm{AB}=15$~m. We choose $G_\mathrm{A} = G_\mathrm{B}=20$~dBi and $G_\mathrm{E} = 25$~dBi. The target semantic security level $\delta$ is depicted for perfect alignment of the transmitter to the position of Bob. Furthermore, in addition to the results depicted in Fig.~\ref{fig_Scenario2_SecrecyMap_dist15}, we fix the transmit power and vary the secrecy code parameter $L$ such that the reliability $\varphi \approx 10^{-3}$ is achieved for horizontal distances between Alice and Bob between $5$~m and $25$~m. For small horizontal distances, the insecure region includes only a small area around the legitimate receiver. With an increasing distance, the size of this area increases. This implies that the maximum horizontal distance $d_{\text{AB}}$ must be limited even for legitimate parties with high-gain antennas to ensure secure communication.
 
\begin{figure}
		\centering
		\input{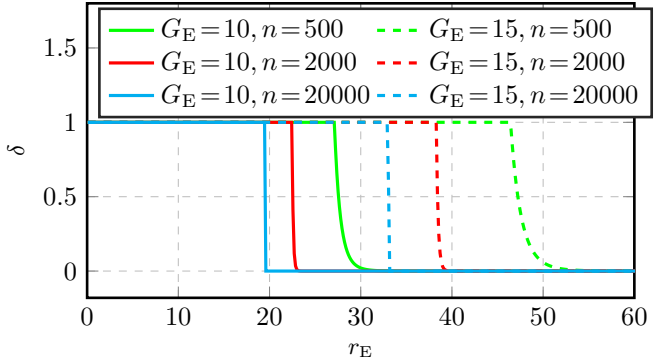}
		\vspace*{-0.4cm}
		\caption{Semantic security level $\delta$ as a function of the horizontal distance $r_\mathrm{E}$ between the transmitter and eavesdropper for multiple blocklengths $n$ and eavesdropper antenna gains $G_\mathrm{E}$. Alice and Bob have the same antenna gains with $G_\mathrm{A} = G_\mathrm{B} = 10$~dBi. Pre-determined parameters: $\varphi \leq 10^{-3}$ and $R = 0.2$. \label{fig_DeltaOverRadius}}
		\vspace*{-0.5cm}
	\end{figure}

	We next analyze the effects of secrecy code parameters on the semantic secrecy levels achieved for Scenario~1. Since the resulting semantic security levels are radially symmetric for Scenario~1, we focus on the horizontal distance $r_\mathrm{E}$ between the eavesdropper and the transmitter. In Fig.~\ref{fig_DeltaOverRadius}, the resulting secrecy levels $\delta(r_\mathrm{E})$, a function of the distance $r_\mathrm{E}$, for multiple blocklengths $n$ are shown. With a larger blocklength, the transition from the insecure region with $\delta = 1$ to the secure region with $\delta = 0$ becomes sharper and a smaller insecure region is observed. For eavesdropper antennas with higher directivity, the area of the insecure region increases. 
	
\begin{figure}[t]
		\centering
%
%
%
\begin{tikzpicture}

\begin{axis}[%
width=\linewidth,
height = 5.5cm,
xmode=log,
xlabel={$\varphi$},
ymax=100,
xmax = 0.11,
xmin = 0.00001,
ylabel={$r_\mathrm{E,0}$},
grid style = dashed,
xmajorgrids, ymajorgrids,
legend columns = 2,
legend pos = north west,
legend style={legend cell align=left, align=left, draw=white!15!black},
label style={font={\small}},
line width=1pt
]

\addplot [color=blue, mark=x, line width=1.2pt, mark size = 3pt,dashdotted, mark options={solid}]
  table[row sep=crcr]{%
0.106508910307776	26.53\\
0.0109352228578108	28.369\\
0.0010171915702407	30.492\\
0.000106068963395988	32.472\\
1.07237112502717e-05	34.736\\
};
\addlegendentry{$G_\mathrm{E} = 10,R = 0.3$}

\addplot [color=blue, mark=o, line width=1.2pt, mark size = 3pt,dashdotted, mark options={solid}]
  table[row sep=crcr]{%
0.106508910307776	45.348\\
0.0109352228578108	48.744\\
0.0010171915702407	52.281\\
0.000106068963395988	55.819\\
1.07237112502717e-05	59.78\\
};
\addlegendentry{$G_\mathrm{E} = 15,R = 0.3$}

\addplot [color=red, dashed, line width=1.2pt, mark size = 3pt,mark=x, mark options={solid}]
table[row sep=crcr]{%
0.102153253280043	21.153\\
0.010836719208461	22.285\\
0.00105356348923926	23.275\\
0.000104545479650853	24.266\\
1.08019064579236e-05	25.398\\
};
\addlegendentry{$G_\mathrm{E} = 10,R = 0.2$}

\addplot [color=red, dashed, line width=1.2pt, mark size = 3pt,mark=o, mark options={solid}]
table[row sep=crcr]{%
0.102153253280043	36.01\\
0.010836719208461	37.849\\
0.00105356348923926	39.689\\
0.000104545479650853	41.528\\
1.08019064579236e-05	43.367\\
};
\addlegendentry{$G_\mathrm{E} = 15,R = 0.2$}

\addplot [color=orange, mark=x, line width=1.2pt, mark size = 3pt,mark options={solid}]
table[row sep=crcr]{%
0.103283931338139	16.908\\
0.0102270308773358	17.474\\
0.00105349531228005	18.04\\
0.000104706648857168	18.606\\
1.01717145713592e-05	19.031\\
};
\addlegendentry{$G_\mathrm{E} = 10,R = 0.1$}

\addplot [color=orange, mark=o, line width=1.2pt, mark size = 3pt, mark options={solid}]
table[row sep=crcr]{%
0.103283931338139	28.511\\
0.0102270308773358	29.643\\
0.00105349531228005	30.492\\
0.000104706648857168	31.341\\
1.01717145713592e-05	32.189\\
};
\addlegendentry{$G_\mathrm{E} = 15,R = 0.1$}
\end{axis}

\end{tikzpicture}%
		\vspace*{-0.8cm}
		\caption{Threshold radius $r_\mathrm{E,0}$ such that $\delta(r_\mathrm{E,0}) = 0.001$ for multiple secrecy code rates $R$ and eavesdropper antenna gains $G_\mathrm{E}$ as a function of the target reliability $\varphi$. Alice and Bob have the same antenna gains with $G_\mathrm{A} = G_\mathrm{B} = 10$~dBi. The secrecy code blocklength is $n = 2000$. \label{fig_r0OverPhi}}
		\vspace*{-0.5cm}
	\end{figure}
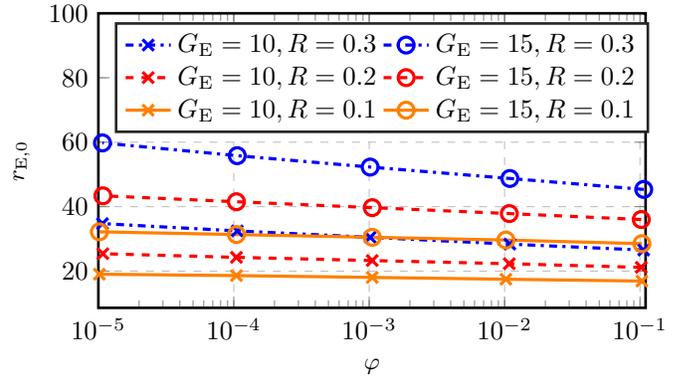

	We next define the threshold radius $r_\mathrm{E,0}$ as the distance between the transmitter and eavesdropper to achieve a semantic secrecy level of $10^{-3}$, i.e., $\delta(r_\mathrm{E,0}) = 10^{-3}$. Fig.~\ref{fig_r0OverPhi} illustrates $r_\mathrm{E,0}$ for multiple secrecy code rates, eavesdropper antenna gains, and target reliabilities. Both a larger code rate $R$ and a larger eavesdropper antenna gain $G_E$ increase $r_\mathrm{E,0}$. Furthermore, if a higher reliability for the legitimate receiver is targeted, corresponding to a smaller $\varphi$ value, $r_\mathrm{E,0}$ increases only slightly for small secrecy code rates.

	\section{Results and Discussions}
	For both presented scenarios, the secrecy maps show a sharp transition between secure and insecure regions for downwards alignment of the transmitter. To ensure secure indoor THz-wireless communication, the access of eavesdroppers to the insecure regions must be prohibited, i.e., the presence of an eavesdropper must be restricted to the secure regions. While weakly directed transmit antennas allow secure communication with multiple legitimate receivers without adaptation of the alignment, the insecure region enlarges as compared to a more directed communication. The insecure region enlarges also for high-gain eavesdropper antennas. 

	With an increasing horizontal distance between the transmitter and legitimate receiver, the region that experiences a high transmit gain enlarges, including a growing area behind the point of alignment. For large distances, the resulting insecure region contains a large portion of the room, including parts of the walls. In this situation, preventing the eavesdropper's access to the insecure region would be even more challenging. Especially the walls could be an easily accessible and hardly detectable point of attack. For example, only a small hole in a wall can be used for inserting an eavesdropping antenna without entering the building. Thus, even for strongly directed transmit antennas, the maximum horizontal distance between the transmitter and the legitimate receivers has to be limited. Moreover, improvements of the THz-wireless communication, such as using multiple antennas, might be necessary to reduce the size of the insecure region.
	
	In practice, the position and antenna gain of an eavesdropper are generally not known. Thus, the worst-case scenarios for the eavesdropper should be considered. An eavesdropper most likely uses strongly directed antennas to overcome the larger path loss and smaller effective transmit antenna gains (due to the transmit antenna misalignment) compared to the legitimate receiver. The reliability, which depends on the secrecy code parameters, and the transmit power should be chosen such that the resulting reliability for the legitimate receiver satisfies a pre-determined target value. The choice of the target reliability does not change the size of the insecure region significantly for the parameters considered. However, the secrecy code blocklength affects the size of the insecure region as well as the sharpness of the transition between the secure and insecure regions. Furthermore, decreasing the secrecy code rate also causes a smaller insecure region. Thus, the strong dependency of the achieved security level on the secrecy code parameters proves the importance of a joint assessment of security in THz-wireless communications taking into account both information-theoretic analysis and methods of communication engineering.

	\section*{Acknowledgment}
	This work was supported by the German Federal Ministry of Education and Research (BMBF) under Grants 16KIS1241K, 16KIS1242, and 16KIS1243.
	
	\bibliographystyle{IEEEtran}
	\bibliography{IEEEabrv,bibliography.bib}
	
\end{document}